\documentclass[sigconf,9pt]{acmart}
\AtBeginDocument{%
  }

\usepackage{enumitem}
\usepackage{subcaption} 
\usepackage{graphicx}
\usepackage{makecell}

\begin{document}

%%
%% The "title" command has an optional parameter,
%% allowing the author to define a "short title" to be used in page headers.
\title[Home Health System Deployment Experience]{Home Health System Deployment Experience for Geriatric Care Remote Monitoring}

% Alzheimer's Disease Patient 
%%
%% The "author" command and its associated commands are used to define
%% the authors and their affiliations.
%% Of note is the shared affiliation of the first two authors, and the
%% "authornote" and "authornotemark" commands
%% used to denote shared contribution to the research.
\author{Dong Yoon Lee}
\affiliation{
    \institution{University of California, Merced}
    \city{Merced}
    \country{United States}}

\author{Alyssa Weakley}
\affiliation{
    \institution{University of California, Davis School Of Medicine}
    \city{Sacramento}
    \country{United States}}

\author{Hui Wei}
\affiliation{
    \institution{University of California, Merced}
    \city{Merced}
    \country{United States}
}

\author{Daniel Cardona}
\affiliation{
    \institution{University of California, Merced}
    \city{Merced}
    \country{United States}
}

\author{Shijia Pan}
\affiliation{
    \institution{University of California, Merced}
    \city{Merced}
    \country{United States}}

%%
%% By default, the full list of authors will be used in the page
%% headers. Often, this list is too long, and will overlap
%% other information printed in the page headers. This command allows
%% the author to define a more concise list
%% of authors' names for this purpose.
\renewcommand{\shortauthors}{Dong et al.}

%%
%% The abstract is a short summary of the work to be presented in the
%% article.
\begin{abstract}
To support aging-in-place, adult children often provide care to their aging parents from a distance. These informal caregivers desire plug-and-play remote care solutions for privacy-preserving continuous monitoring that enabling real-time activity monitoring and intuitive, actionable information. This short paper presents insights from three iterations of deployment experience for remote monitoring system and the iterative improvement in hardware, modeling, and user interface guided by the Geriatric 4Ms framework (matters most, mentation, mobility, and medication). An LLM-assisted solution is developed to balance user experience (privacy-preserving, plug-and-play) and system performance.
\end{abstract}

%%
%% The code below is generated by the tool at http://dl.acm.org/ccs.cfm.
%% Please copy and paste the code instead of the example below.
%%

% \ccsdesc[500]{Do Not Use This Code~Generate the Correct Terms for Your Paper}
% \ccsdesc[300]{Do Not Use This Code~Generate the Correct Terms for Your Paper}
% \ccsdesc{Do Not Use This Code~Generate the Correct Terms for Your Paper}
% \ccsdesc[100]{Do Not Use This Code~Generate the Correct Terms for Your Paper}

%%
%% Keywords. The author(s) should pick words that accurately describe
%% the work being presented. Separate the keywords with commas.
%% A "teaser" image appears between the author and affiliation
%% information and the body of the document, and typically spans the
%% page.
% \begin{teaserfigure}
%   \includegraphics[width=\textwidth]{sampleteaser}
%   \caption{Seattle Mariners at Spring Training, 2010.}
%   \Description{Enjoying the baseball game from the third-base
%   seats. Ichiro Suzuki preparing to bat.}
%   \label{fig:teaser}
% \end{teaserfigure}

% \received{20 February 2007}
% \received[revised]{12 March 2009}
% \received[accepted]{5 June 2009}

%%
%% This command processes the author and affiliation and title
%% information and builds the first part of the formatted document.
\maketitle

\section{Introduction}
% \hw{if we only targeting at AD patients, we can say that directly here, instead of saying old people first and then use an example to introduce AD patients. We can say something like Here are many AD patients, which gets physical, emotional, and financial burden to patients, their families and the whole society. Although there are so many AD patients, there is a critical caregiver shortage in the U.S., which cannot take care of these AD patients, and ......}
% \sj{let's keep it general and AD is addressed as mentation consideration}
The United States faces a critical caregiver shortage, a challenge exacerbated by the rapidly growing older population. In 2020, 55.8 million individuals (17\% of the population) were aged 65 or older \cite{olderPopulation}, a number projected to rise to 85.7 million (23\%) by 2050 \cite{olderPopulation2}. 
Aging is associated with a rise in health conditions requiring medical management and a decline in cognitive and physical conditions which threatens a person to live independently. To assist older adults age-in-place, friends and family, mainly adult children, often take on caregiving roles from a distance. Family members who take on informal caregiving roles contribute an estimated \$600 billion annually in unpaid support \cite{olderPopulation4}. These informal caregivers desire digital tools to assist them in remote care delivery, monitoring, and actionable insights. 

Technology advancements in the Internet of Things (IoTs) and artificial intelligence (AI) have the potential to improve informal caregiving as well as older adult quality-of-life by enabling real-time activity monitoring \cite{Zhou}, just-in-time support \cite{Muller2017Sep}, and actionable information \cite{Leslie2021Mar} for caregiver and clinical support teams. 
% \hw{target this sentence at AD patients instead of general old people.} \sj{I agree, here would be a good place to mention older adult challenges and missing from prior work}
For example, prior work has demonstrated the feasibility of capturing and recognizing activities under the lab environment with healthy young adults using sensors such as mmWave \cite{Singh2019Oct}, vibration \cite{Gordon}, PIR \cite{Kashimoto}, and wearables \cite{Lara2012Nov}.
Corresponding machine learning algorithms have been developed to predict target information for different people in varying environments, mitigating the impact of domain variances \cite{Chen2018Mar, Shi2022Jul}.
% Many prior studies have focused on developing machine learning techniques for remote monitoring \cite{}; however, hardware- and user-related c
However, challenges in real-world deployments with older adults and their caregivers remain largely unaddressed. 
% These challenges, together with the limitations of machine learning models, create significant barriers to the adoption of remote monitoring solutions by families of Alzheimer’s disease patients. 
For example, limited domain knowledge about sensor systems and their configurations across diverse deployment environments may hinder families and caregivers from effectively using home-based monitoring systems. 
Suboptimal deployment and configuration  can result in insufficient or low-quality data, reducing the model's ability to profile users accurately. 
Declining memory and other aspects of cognition may impact the usability and feasibility of adopting technology. 
% In addition, machine learning models often face computational constraints when deployed on edge devices, which can further degrade monitoring performance. 
Without bridging the gaps among hardware, modeling, and end users, these challenges will continue to impede the practical adoption of remote monitoring technologies for geriatric care at home.

%Many of these prior works focus on the machine learning for remote monitoring, but the other hardware and user related challenges for real world deployments are yet to be addressed.

% \color{purple}
% Certain challenges especially make adoption for these remote monitoring solutions not possible for AD patient families. \hw{Change this sentence to "However, these challenges combined with challenges of machine learning pose significant barriers to the adoption of remote monitoring solutions by families of Alzheimer’s disease patients."}
% For example, implicit domain knowledge for sensor systems and different deployment environments may lower the user experience for family members of AD patients to configure their own remote monitoring system from in-home use. \hw{Change this sentence into "For instance, lack of implicit domain knowledge of sensor systems and their configuration in varying deployment environments may hinder the usability of home-based remote monitoring systems for families of Alzheimer’s disease patients."}
% If the hardware is misconfigured, there may be insufficient data to sufficiently monitoring users.
% Finally, machine learning models may have limited computing resources for inference on edge devices, which may hurt monitoring performance.
% Without addressing these gaps between end users, system hardware, and machine learning, these challenges limit adoption of these remote monitoring solutions for AD patients.

To fill in these gaps,
we present a plug-and-play long-term continuous monitoring solution with ambient vibration-based sensing system. 
We adapt multiple aspects of the system based on the \textbf{Geriatrics 4Ms Framework} -- (what) matters most, mentation (cognition and mood), mobility, medication -- to systematically reflect the needs of older adults and their caregivers.
We identify and tackle challenges in hardware (Section \ref{sec:hardware}), (data) modeling and inference (Section \ref{sec:model}), and user interface (Section \ref{sec:user}), to enable (1) plug-and-play deployment, (2) robust on-edge inference, and (3)user-friendly experience for older adults with cognitive decline and their family caregivers.

First, we develop product-grade sensing device hardware and customize network protocol and configuration based on the individual home deployment to enable \textit{plug-and-play deployment} for end users.
We conduct semi-controlled experiment in a physical simulation suite, which is equipped with surveillance video and microphone for ground truth collection.
In this deployment, we generate a dataset with multiple participants and second-level ground truth label.
The unexpected network issues, including congestion, limiting access to the Internet, and unknown sensing device status, are addressed with updated network design before the in-home uncontrolled data collection.

Second, we establish data collection protocol to ensure sensing quality and efficient deployment that are contextual relevant.
To ensure data security and privacy, we develop robust activity recognition models that can be deployed \textit{on edge} devices. 
A temporal convolutional network (TCN) model for vibration signals is fine-tuned from a pre-trained model originally trained on online acoustic data.

Finally, we develop an LLM-assisted interactive deployment recommendation system to facilitate intuitive deployment and usage for non-expert end users.
LLM's broad world knowledge and strong natural language understanding capabilities allows effective reasoning for deployment configuration recommendation. 
However, technical gaps remain in enabling LLMs to comprehend the spatial aspects of IoT deployments and the implicit requirements of users (e.g., cognitive decline), which are areas that prior work has largely overlooked. 
In particular, current approaches lack methods for generating consistent and physically feasible recommendations that \textit{balance system performance with user experience}.
We tackle this by designing schema templates to supply the LLM with a structured representation of the sensing system, environment, and user preferences.
Then we use instruction prompting to turn the user preferences into a recommendation for the sensor configuration.

In summary, the contributions of our paper are as follows:
\setlist{nolistsep}
\begin{enumerate}[noitemsep,leftmargin=*]
    \item We present a home health monitoring system that is customized for older adults and their caregivers based on Geriatric 4Ms framework.
    \item We address the hardware challenges for plug-and-play real-world deployments, including a custom fault-tolerant protocol and network design.
    % \item We address these challenges with a custom fault-tolerant protocol and network design.
    \item We establish the deployment protocol to ensure sensing quality and data relevancy, and develop robust learning models for activity recognition that can be deployed on edge devices. 
    \item We design an LLM-powered recommendation system, outputting a user-tailored deployment protocol that balances system performance and user experience.
    \item We conduct multiple rounds of system evaluation and user study with real-world deployments to demonstrate our system's effectiveness.
\end{enumerate}

% The rest of the paper is organized as follows. First, Section \ref{sec:system} introduces the iterative design and deployment process of the system. Then, we further analyze the key impact factors on the performance and user experience of the system in Section \ref{}.

\section{Geriatrics 4Ms Framework Guided Home Health System Design}
The Geriatrics 4Ms framework -- (what) matters most, mentation, mobility, and medications \cite{G4Ms, mate2021evidence} -- was developed as a guideline for the design of care or services for older adults. 
We leverage the 4Ms framework to adjust the system and deployment designs.
% \subsubsection{(What) Matters Most.} For both patients and their adult children caregivers, they mentioned their 
% \subsubsection{Mentation.}

\subsection{Plug-and-Play Hardware and Network Design}\label{sec:hardware}
The goal of hardware design is stable and scalable data collection in home environments. This section introduces the sensing device and network designs to achieve this goal and how the 4Ms framework guides our design.

\begin{figure}[!t]
    \centering
    \includegraphics[width=\linewidth]{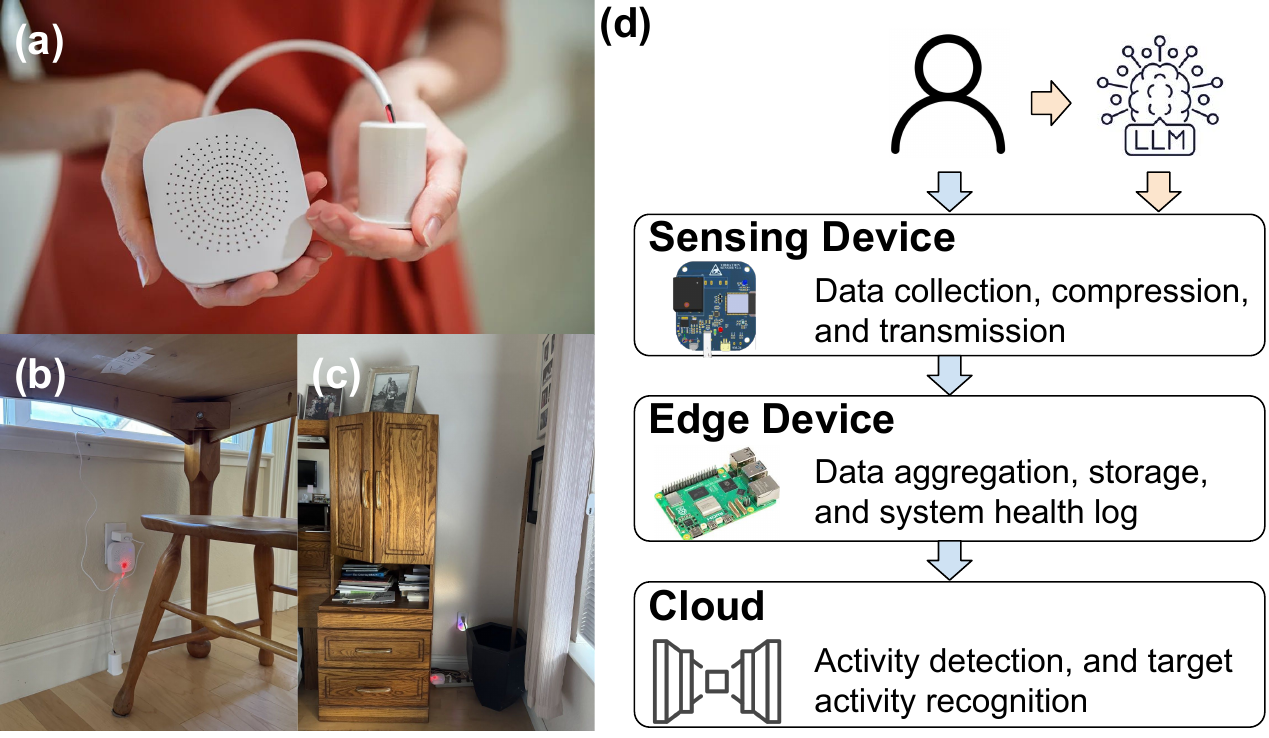}
    \caption{Plug-and-play system for home health monitoring. (a) depicts the plug-and-play sensing device connected to a vibration sensor. (b) and (c) show the system working in a real-world deployment. (d) presents the system design for LLM-assisted deployment, where the blue arrows indicate sensing data and the orange arrows show deployment data.}
    \label{fig:intuition}
\end{figure}

\begin{figure*}[!t]
  \centering
  \begin{subfigure}{0.33\textwidth}
    \centering
    \includegraphics[width=\linewidth]{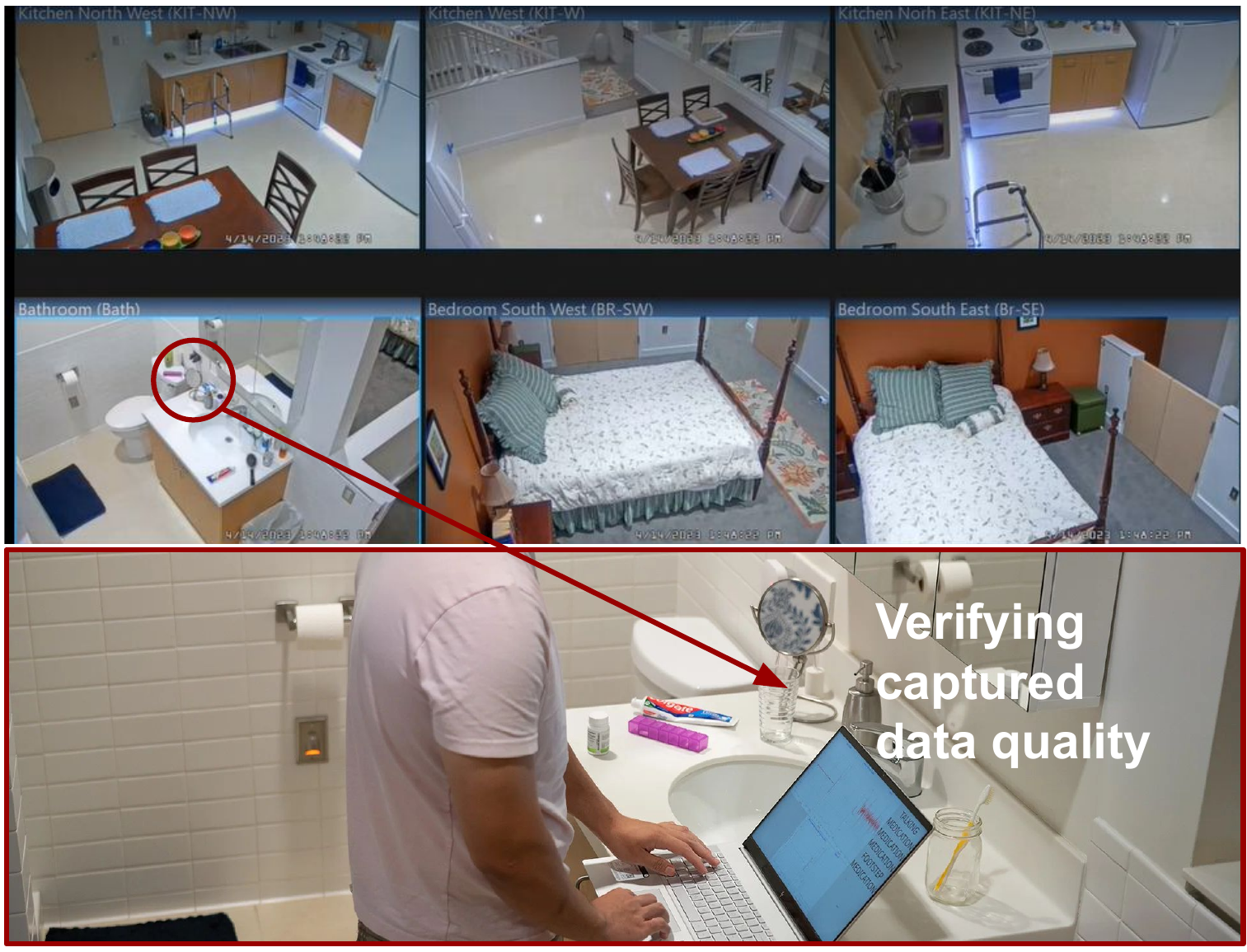}
    \caption{Deployment 1: Simulation Suite.}
    \label{fig:dep1}
  \end{subfigure}\hfill
  \begin{subfigure}{0.33\textwidth}
    \centering
    \includegraphics[width=\linewidth]{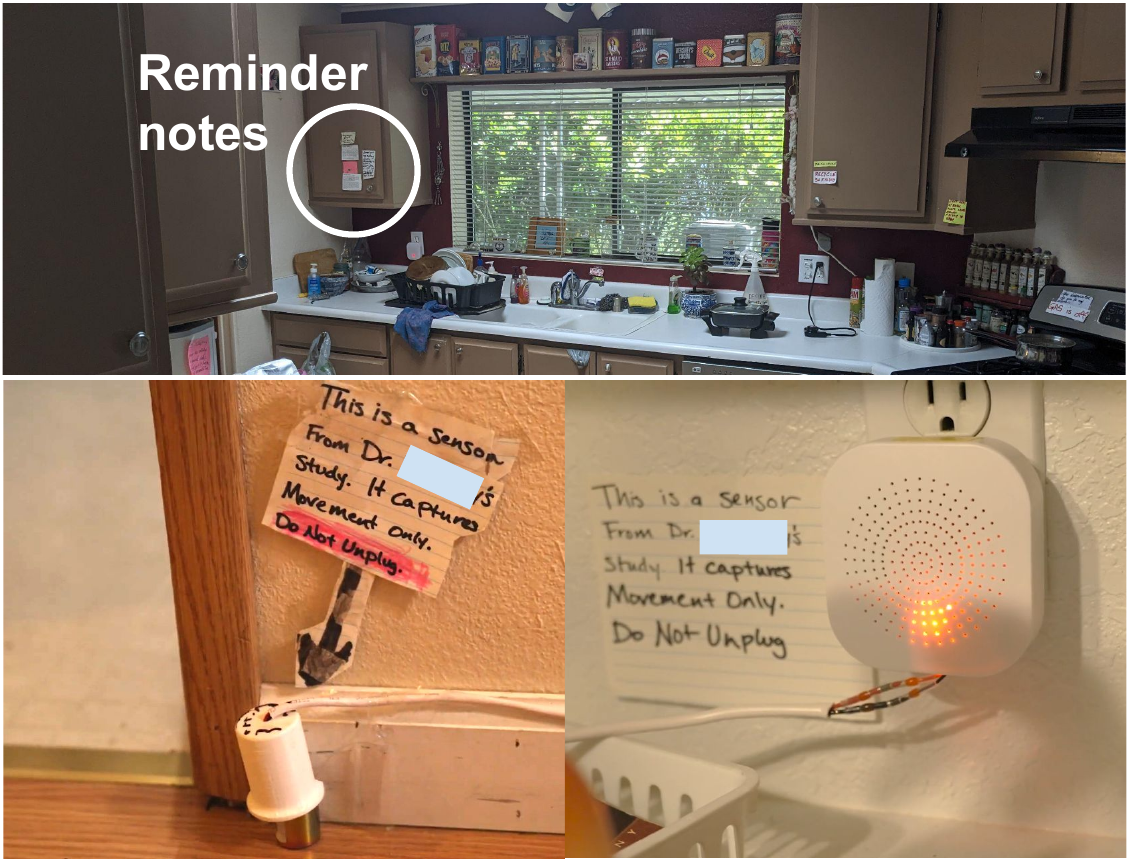}
    \caption{Deployment 2: Participant \#1.}
    \label{fig:dep2}
  \end{subfigure}\hfill
  \begin{subfigure}{0.32\textwidth}
    \centering
    \includegraphics[width=\linewidth]{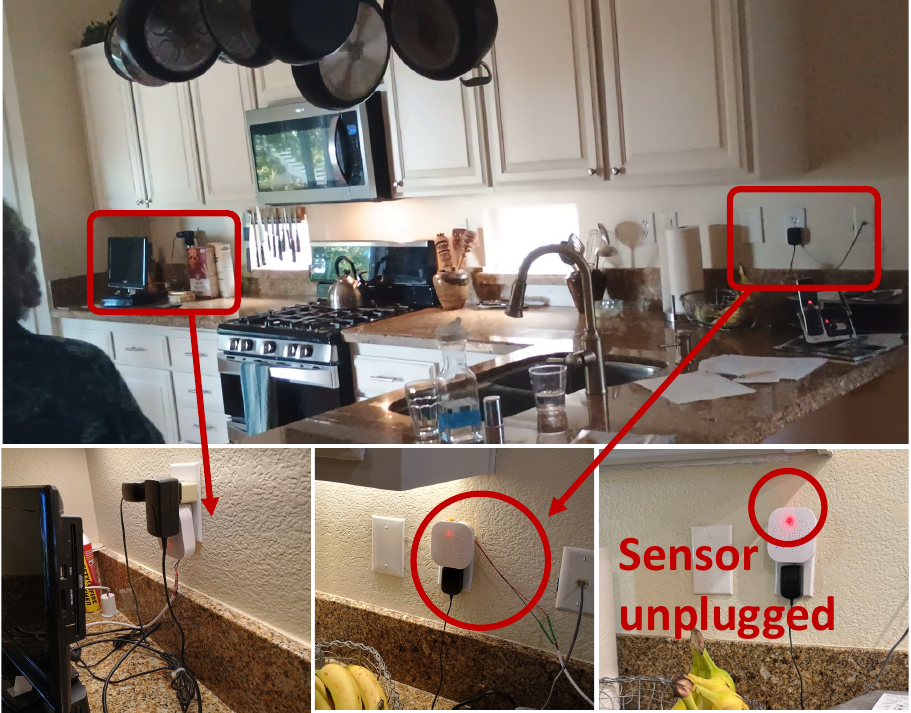}
    \caption{Deployment 3: Participant \#2.}
    \label{fig:dep3}
  \end{subfigure}
    \vspace{-2ex}
  \caption{Three iterations of deployments.}
  \Description{Brief text-only description for accessibility (required by ACM).}
  \label{fig:three-across}
\end{figure*}

\subsubsection{Plug-and-Play Device Design}
The key challenge for the sensing device design is to be as discreet as possible due to the following reasons.
First, many older adults do not want their visitors to notice these assistive devices, as they value their sense of independence and privacy (what matters most). 
Second, older adults with cognitive impairment may tamper with a device they are unfamiliar with (mentation).
Third, consistent with general design principles, people appreciate devices that are aesthetically pleasing and/or blend in with surroundings. 
To meet these needs, we iterated three times in a co-design process with an expert in AgeTech design and older adults with cognitive impairment. For example, our initial printed circuit board design was replaced with a white product-grade enclosure with the wall-plug design, as shown in Figure \ref{fig:intuition}, to enhance discretion and acceptance. We also deemed it essential to cover the sensor and wire properly for cohesion and to prevent unnecessary attention, which could lead to tampering with the device.

The device consists of an \textit{ESP32S3-Mini-1-N8} processor, equipped with an \textit{ADS131M02} ADC module. 
The sensor used to capture vibration is a geophone, which is a cylinder-shaped metal device in contact with the monitoring surface.
The human-surface (mobility) and object-surface (medication) contacts can generate vibration signals in the range of 10 Hz to 1000 Hz. 
The system samples the geophone sensor at around 7000 Hz to ensure the efficient capture of these signals and to avoid high frequency noise aliasing.

% Because AD patients may forget about the deployment or the use of the sensor, it is important to appropriately cover the sensor and wire, as shown in Figure \ref{fig:intuition} (b). 
% Otherwise, the exposed wires and the sensor may draw unnecessary attention, resulting in tampering of the device.

\subsubsection{Plug-and-Play Network Design}
Each home was monitored by four to five sensing devices, which communicate wirelessly with a local edge device.
The edge device serves as the in-home hub for data storage, lightweight signal processing, and device health monitoring (e.g., system faults, connectivity). 
% In the future, the inference will also be done on the edge device. 
For research purposes, we store the data on an external drive connected to the edge device and manually retrieve it on a weekly basis to ensure the privacy and security of the user data.
Since each device samples the sensor at around 6800 Hz, it generates around 16 GB data per day. 
We observe that directly streaming data to the edge causes network congestion and packet errors. 
Therefore, we modified the User Datagram Protocol (UDP) layer to lessen overhead. 
We further implemented packet-level parity protection to detect and correct corrupted packets.
If a packet contains a bit-level error, we recover the information using the parity information sent with the packet.

To facilitate smooth deployment at end-users' homes, we pre-configure the network connection between the sensing devices and the edge device. 
The user only needs to connect the edge device to their home router to enable remote system status updates.
To accommodate heterogeneous firewall policies, we established an authorized tunneling mechanism to ensure consistent connectivity across sites.
We use \textit{wireguard} to create a direct tunnel from the edge device to the cloud, without any port forwarding or home network configuration necessary, such that we can send over the overall system health, real-time updates on system performance, and hard drive space availability.

\subsection{Robust Modeling to Address Sensing Quality and Scenario Variance }\label{sec:model}
The goal of the modeling of this system is to recognize key activities that fall within the framework of the Geriatric 4 Ms and are of interest to the clinicians, caregivers, and individuals with cognitive impairment, including self-care, taking \textit{medication}, and \textit{mobility}. 
There are two major system modeling challenges.
First, to ensure the designated information is captured, the sensors need to be placed strategically at spatially relevant locations (what \textit{matters most}).
Second, to allow effective prediction at different sites, the model needs to be robust to domain variances. 
% posed to the problem: (1) because the system adopts general-purpose sensing (vibration), it is important to ensure the system can effectively capture 

\subsubsection{Deployment Protocol Prioritizing Data Quality}
Compared to other sensing systems that require line-of-sight to conduct sensing, vibration-based sensing allows for more flexible deployment by simply plugging into the outlets in the environment. 
However, as a result, the power outlet locations constrain the sensor placement, resulting in varying performance for target activity detection and recognition.  
When deploying sensors, we ask participants to conduct a list of scripted activities and verify on-site whether the deployed sensors can capture the induced vibration signals of the target activities. 
The domain expert will visually inspect the signal to verify there is no significant problem with the data capture.

\begin{figure}[!t]
    \centering
    \includegraphics[width=\linewidth]{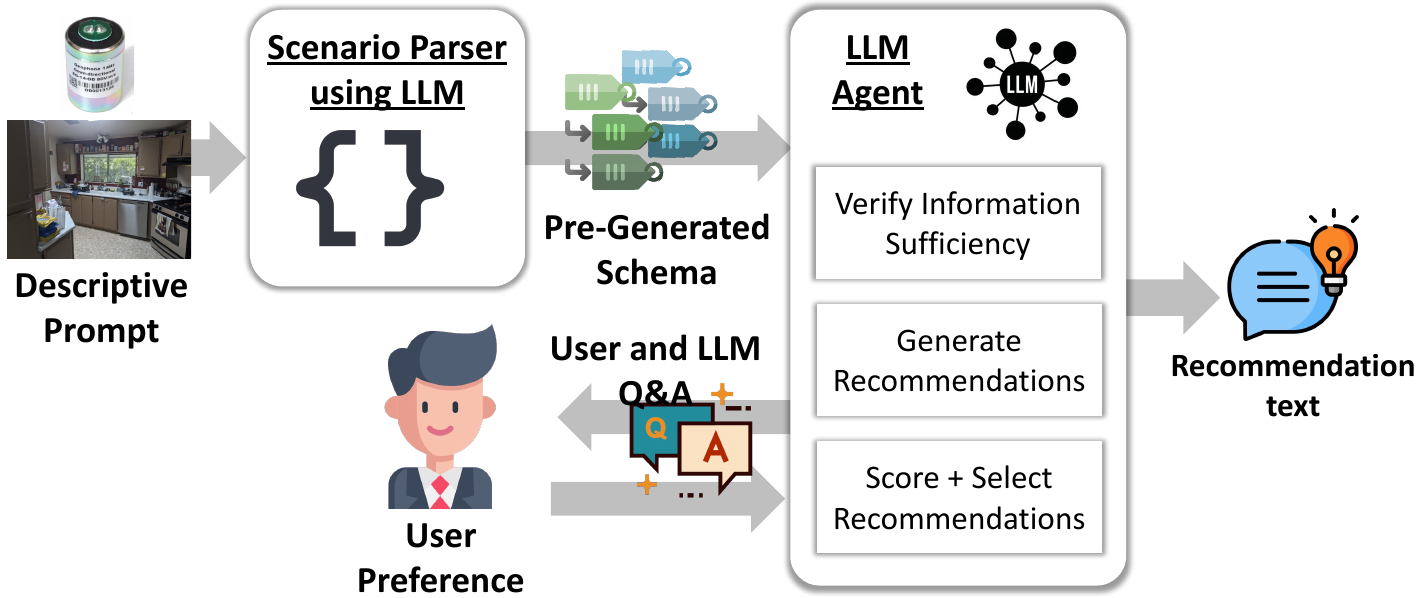}
    \vspace{-2ex}
    \caption{LLM-assisted deployment recommendation system.}
    \vspace{-1em}
    \label{fig:llm}
\end{figure}

\begin{table*}[!t]
    \centering
     \caption{Dataset Overview of the 3 Deployments}
     \vspace{-1.5ex}
    \begin{tabular}{|c|c|c|c|c|c|c|c|}
    \hline
     Deployment & Focus & Participants & Location & Label  & Networking & User Interface & Storage \\ \hline \hline
     1 & system & 4 healthy young adults& simulated apt.&  120 mins &  UDP & None & 150GB/day\\
     2 & user & 1 AD patient (84 y/o) &  home \#1 & 0 min &  UDP w/ parity & None & 150GB/day\\
     3(B) & sys+user & 1 AD patient (88 y/o) & home \#2 & 30 min &  UDP  w/ parity & None & 70GB/day\\
     3(A) & sys+user & 1 AD patient (88 y/o) & home \#2 & 30 min  &  UDP  w/ parity & \makecell{LLM agent} & 70GB/day\\
     \hline
\end{tabular}
    % TODO: put info column in section Dataset Comparison
    \label{tab:dataset_overview}
\end{table*}

\begin{figure*}[!t]
    \begin{subfigure}{0.3\linewidth}
        \centering
        \includegraphics[width=\linewidth]{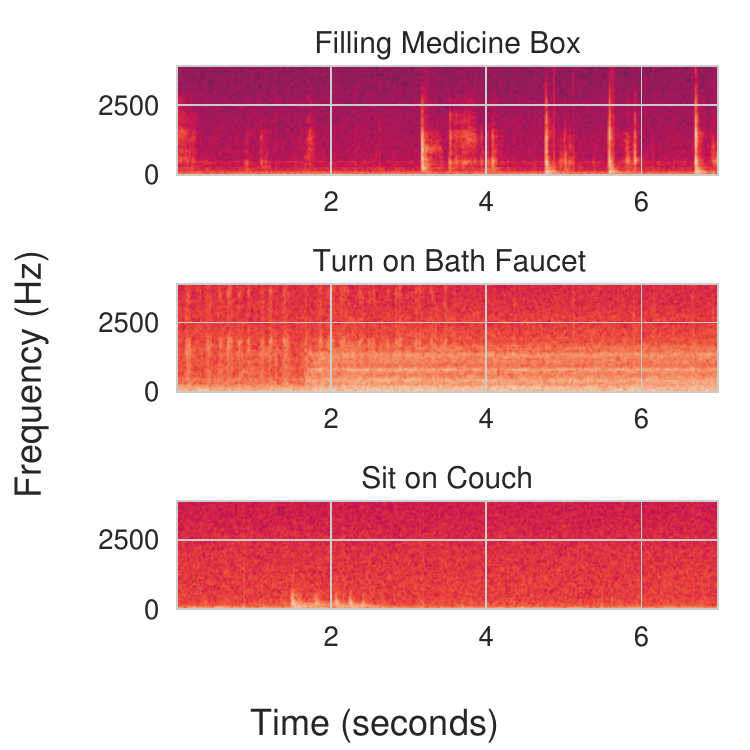}
        \caption{}
        \label{fig:event_sig}
    \end{subfigure}
       \begin{subfigure}{0.6\linewidth}
        \centering
        \includegraphics[width=\linewidth]{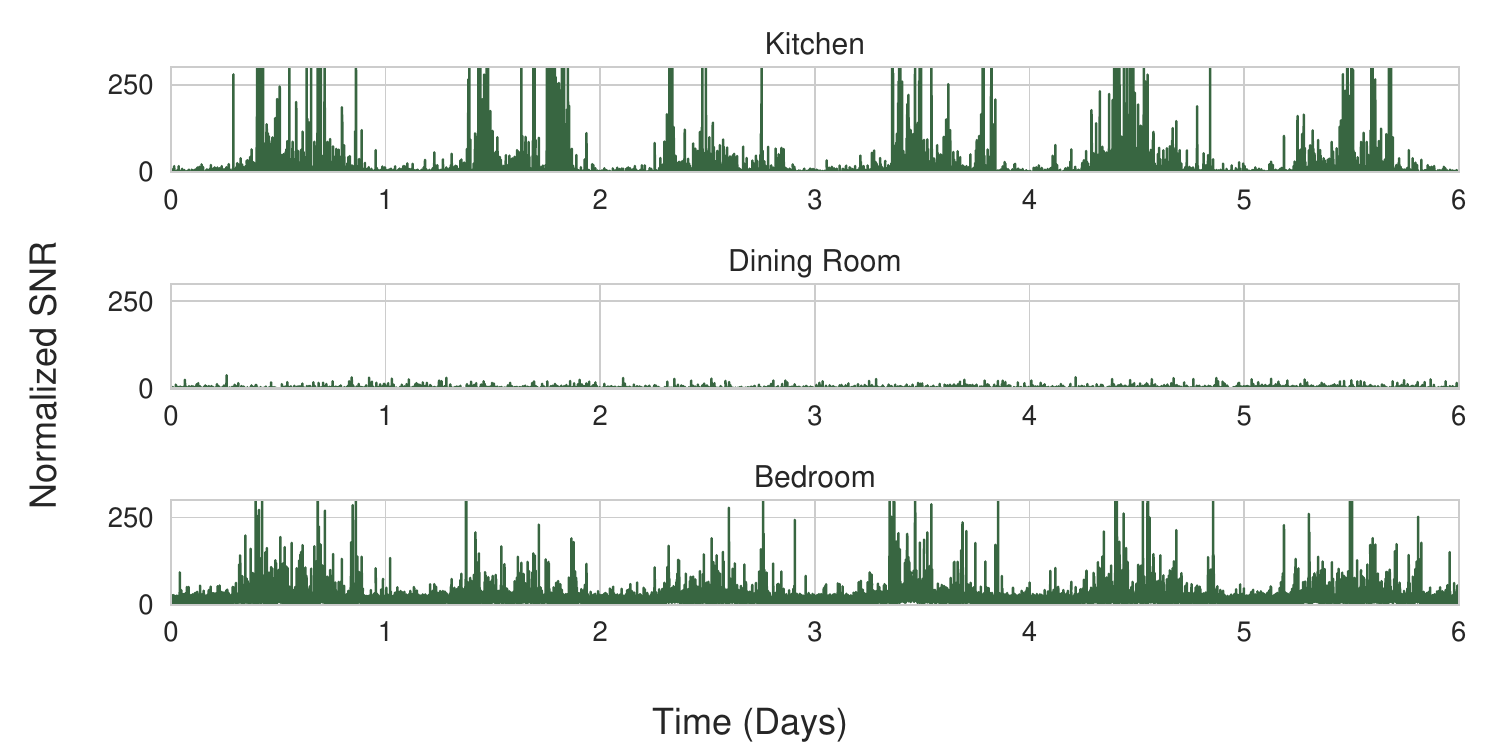}
        \caption{}
        \label{fig:week_sig}
    \end{subfigure} 
    \vspace{-2ex}
    \caption{Example signals. (a) Spectrogram of vibration signals from Deployment 1. Three labeled activities, including filling a medicine box, turn on the shower, and sitting on the couch are shown different time-frequency characteristics. (b) Normalized SNR of four sensors in Deployment 3 for 1 week. We observe repeating daily activity patterns in each room.}
    \label{fig:signals}
\end{figure*}

\subsubsection{Activity Recognition Model}
The major challenge for modeling activities is the data domain shifts between different deployments, such as users and environments variances.  
To enable robust modeling across multiple sites, we follow prior work on vibration-based classification models \cite{kimura2024vibrofm, chang2025leveraging, lee2025rarr} and developed a multitask learning model with an unsupervised signal reconstruction task and a supervised learning task using audio data online. 
The model is then fine-tuned with a small amount of vibration signals for activity recognition.
Note that we include the machine learning model here for the completeness of the system design, without claiming the technical contribution here.

To ensure the model can be run on the edge device for inference, we design the model architecture based on temporal convolutional neural networks (TCN)  \cite{bai2018empirical} to encode the data into a smaller latent representation. 
By classifying with only the latent encoding instead of the raw signal, we obtain efficient inference performance on the edge device.

\subsection{LLM-Assisted Deployment to Balance Performance and User Experience}\label{sec:user}
A user-friendly plug-and-play solution is desired to improve remote monitoring ease and accessibility. Therefore, a primary objective is to create a system that is effectively and independently deployed by non-expert end users (i.e., family caregivers).
However, end users may not know how the sensing system works and therefore do not have the knowledge to make effective decisions on the sensor placement to satisfy monitoring needs.

% \subsubsection{LLM-Assisted IoT Deployment Recommendation}To effectively include (what) matters most and the mental conditions, which are often described with natural language, we developed a Large Language Model (LLM) tool that takes both (what) matters most and the mental condition as natural language inputs for system deployment recommendation.
To make recommendations taking both sensing expert knowledge, in-situ environment, and user preference context into account, we leverage LLM's reasoning capability.
To allow consistent and physically feasible reasoning, we adopt a structured data representation with graph-based schemas to organize the sensor and environment information.
Figure \ref{fig:llm} depicts our design, where text descriptions of the environment and sensors is filled in a schema template in order to generate the structured information.
% In order to generate these schemas, a parser, shown in Figure \ref{fig:llm}, uses text descriptions of the environment and sensors to fill in a schema template in order to generate the structured information.
% With these schemas as an input, the LLM requires a way to process the knowledge with reasoning that embeds expert domain knowledge and the user's preferences in order to generate a recommendation that is acceptable for the user and is sufficient for remote monitoring.
We design an expert LLM agent, using chain of thought reasoning \cite{Yu2023Oct}, to reason like a domain expert in sensor systems performing 3 steps -- get sufficient information from the user, generate a list of recommendations, score and select these recommendations.
The user preference, such as privacy concern, cognitive decline (may lead to sensor tampering), is collected by tasking the LLM to ask family members or users questions.
% In order to get any other information about the deployment, such as if the user has a privacy concern, is more prone to tamper, or more information from the environment, the LLM is tasked to ask family members or users questions about the user's preferences, in order to accurate predict the user experience for a given recommendation.
Once having sufficient information, the LLM generate a list of recommendations for sensor configurations, such as location, orientation, and gain.
The LLM then scores each recommendation by sensing performance using the expert reasoning from the chain of thought prompting, and predicts the user experience as a score, using the answers the user has given.
The LLM uses these two metrics to sort the recommendations and choose the best recommendation based on the sum of these metrics, equally weighted.
In this way, we allow users to expertly control their system without any knowledge requirements by using an expert LLM agent.

\begin{figure*}
\begin{subfigure}{0.30\linewidth}
    \centering
    \includegraphics[width=\linewidth]{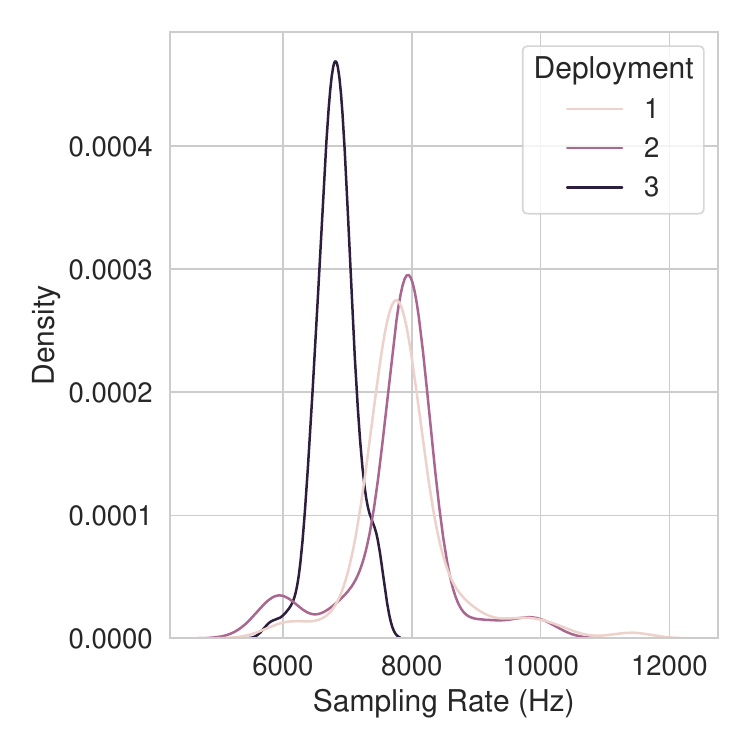}
    \vspace{-1.5ex}
    \caption{}
    \label{fig:sampling-rate}
\end{subfigure}
\begin{subfigure}{0.30\linewidth}
    \centering
    \includegraphics[width=\linewidth]{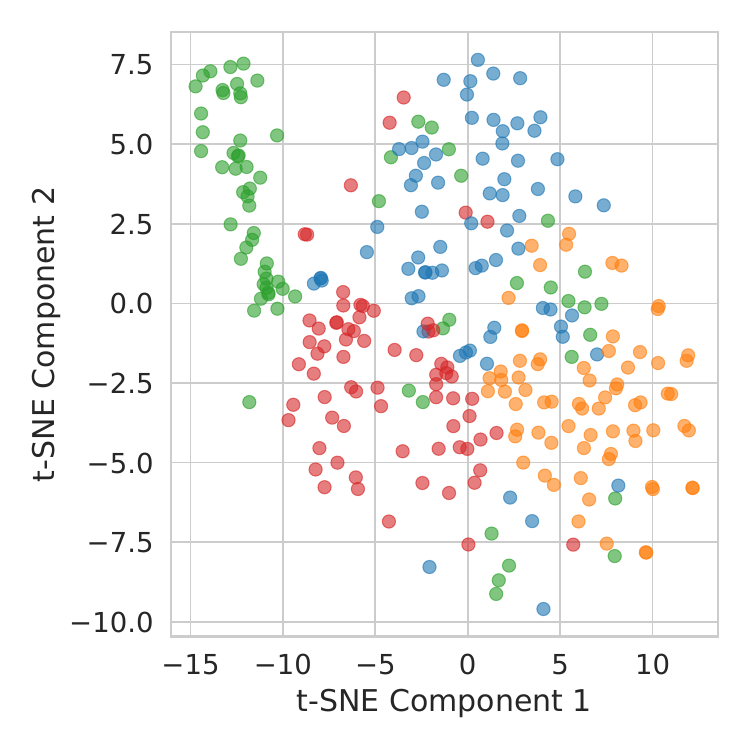}
    \vspace{-1.5ex}
\caption{}
    \label{fig:tSNE}
\end{subfigure}
\begin{subfigure}{0.30\linewidth}
    \centering
    \includegraphics[width=\linewidth]{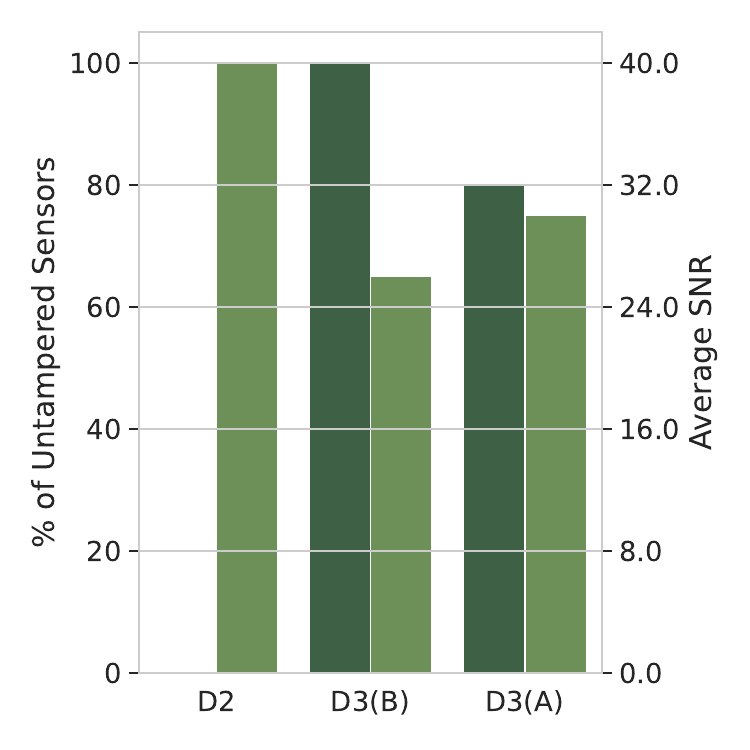}
    \vspace{-1.5ex}
\caption{ }
    \label{fig:efficiency}
\end{subfigure}
\vspace{-2ex}
    \caption{Quantitative analysis. (a) Deployment hardware comparison. The sampling rate in Deployment 1,2,3 are presented by the light pink, red, and black solid line. (b) t-SNE plot for activities of putting a cup/object down (blue), turning on the shower (yellow), walking (green), refilling a medicine bottle (red). (c) Deployment efficiency comparison. The dark green bars depict percentage of untampered sensors, and the light green bars present average SNR of the activity signals. }
    \label{fig:quant}
\end{figure*}

\section{Evaluation and Analysis}
All deployments and data collection procedures adhered to the approved Institutional Review Board (IRB) protocol.

\subsection{Deployments, Datasets, and Labeling}
We conduct three iterations of deployments. These deployments generate the datasets summarized in Table \ref{tab:dataset_overview}. 
% The iterative improvements of the deployment are shown quantitatively.

\subsubsection{Deployment 1: Simulation Suite for Semi-Controlled Data Collection}
First, we conduct a pilot deployment in a Home Health Simulation Suite, a fully functioning 1-bedroom apartment equipped with video cameras and microphones for ground truth collection. 
This U-shaped simulation suite is connected to an observation room, which allows examiners to view activities throughout the apartment, as shown in Figure \ref{fig:dep1}. The suite also includes an intercom system for communication with participants.

We collected data from four participants over a 7-day deployment period. 
These participants were healthy young adult research volunteers. The participants were asked to complete a variety of sequenced activities, such as walking from room to room, brushing their teeth, taking medication from a pill bottle or pill box, and opening or closing doors.

Because the Simulation Suite does not have public networks for data security reasons, we synchronize the clocks between the ground-truth camera and the edge device using a custom-designed portal.
We conduct manual ground truth labeling at the second level for 2 hours with the camera data. 
We then align the detected vibration event signals with the camera ground truth.
Figure \ref{fig:event_sig} depicts signal examples in the form of spectrogram for three types of activities that clinicians are interested in -- medication interaction, taking bath, and sitting on couch.

\subsubsection{Deployment 2: Real-World Home Data Collection \#1}
Based on Deployment 1, we conducted Deployment 2 in the home of an 84-year-old non-Hispanic White female with a clinical diagnosis of dementia (Montreal Cognitive Assessments scores of 20/30) who was living alone in the community. 
It was observed that the participant and family utilize sticky notes as a form of memory compensatory strategy, as shown in Figure \ref{fig:dep2}.
Thus, we adopted the same method to remind the user what the sensor was and to not unplug the sensor by placing a note next to the sensor.
Unfortunately, this strategy did not prevent the participant from tampering with and unplugging the devices.
We conduct a week-long data collection, documenting the system health and user experience of having in-home sensor systems for remote monitoring.
Due to their memory impairment, during the post-study interview, the participant was unable to recall removing the device.
This further confirms the need for discreet deployment and user-friendly design. 

\subsubsection{Deployment 3: Real-World Home Data Collection \#2}
Leveraging results from Deployment 2, we conducted Deployment 3 in the home of an 88-year-old Hispanic White female with a clinical diagnosis of dementia (Montreal Cognitive Assessments scores of 19/30) who was living alone in the community. 
Figure \ref{fig:dep3} depicts the deployment environment.
Two sets of sensing systems (each with 5 sensors) are deployed at 5 areas, which are referred to as Deployments 3(A) and 3(B) in Table \ref{tab:dataset_overview}. 3(A)'s sensor placement is determined by the LLM with prompts to indicate environment and user information. 3(B)'s sensor placement is determined by the patient's son, who is one of the major remote caregiver. 

We conducted a week-long data collection, documenting system stability, as well as user experience.
Figure \ref{fig:week_sig} showcase normalized SNR for three sensors placed in the kitchen, dining room, and bedroom, respectively.
We can observe daily activity patterns, variation, and transition between rooms.
At the end of the week-long deployment, we captured 30 minutes of ground truth video data for labeling.
In those 30 minutes, we asked the participant to demonstrate their normal daily activities they would carry out in various parts of the home such as opening/closing doors/cabinets, using the computer, brushing their teeth, taking medication, and lying on their bed.
Using the system health monitoring tool we developed, as well as our observation of the sensor units after the one-week period, we also analyzed if there was any user tampering with the sensor units.

\subsection{Quantitative Analysis for System Performance}
We further demonstrate a detailed analysis of system performance in Figure \ref{fig:quant}.
Figure \ref{fig:quant} (a) shows the sampling rate distributions of each packet over three deployments.
% the data quality, including sampling rate, separability of signal features, tampering rate, and signal-to-noise ratio (SNR) of the detected signals in Figure \ref{fig:quant}.
% In this section, we analyze the quantitative measurements of each hardware configuration, potential ML performance for activity recognition, and physical configuration. 
% The sampling rate, where Deployment 2 and 3 demonstrates \textcolor{red}{xxx}.
% In order to see the performance of different hardware configurations, we compare the sampling rate distributions of each packet for each deployment in Figure \ref{fig:sampling-rate}.
We can see that Deployment 3 shows less variance, with a standard deviation of $316$ Hz, in Figure \ref{fig:sampling-rate}, compared to Deployments 1 and 2, each with standard variance of around $734$ Hz and $799$ Hz, respectively.
% This is the result of the added parity information to each packet.
% However, Figure \ref{fig:sampling-rate} also shows that it lessens the average sampling rate to $6800$ Hz, rather than around $7800$ Hz for Deployment 1 and 2.
It demonstrates that optimizing sampling rate to synchronize between edge device and sensor unit results in a more consistent sampling quality.

To understand how effective is the data collected can be used for activity recognition tasks, we investigate four activities of interest: putting down objects, turning on the shower, walking, and interaction with a medicine bottle.
We set the perplexity to $n=30$, and use the FFT of the activity signals as input features for the t-SNE plot.
Fine-grained labeled data from Deployment 1 is used to illustrate the feasibility here with a t-SNE plot (KL Divergence  $= 0.7438$), where 284 activities from 4 classes are visualized in Figure \ref{fig:tSNE} with separable feature space.
% From the 4 classes of 284 activities from Deployment 1 in Figure \ref{fig:tSNE}, they show to be visually separable.
% The previously stated configuration are used to fit these activities to give a KL Divergence of $0.7438$.
This t-SNE plot verifies that our system can effectively capture signals of context relevant event for older adults remote monitoring.
% the features from the configured sensors, and our deployment method from Deployment 1, is sufficient for human activity recognition and other related tasks to remote monitoring.

Finally, we quantify the effectiveness of the LLM recommendation system in Figure \ref{fig:efficiency} by comparing the percentage of sensors untampered by the user and the average SNR.
% balances the performance of the sensor system as well as reduce user tampering of sensor systems.
We calculate SNR as a ratio of all the signal energy over signal energy of ambient environmental noise, and normalize this ratio based on the time duration.
We define the percentage of untampered sensors as the ratio between the number of sensors remaining untampered after 1 week and the total number of sensors initially deployed.
The percentage of untampered sensors of Deployment 2, 3(B), and 3(A) are shown as dark green bars.
We observe that Deployment 2, which prioritizing data quality, gets the highest average SNR. However, the sensor locations are not efficient to avoid user tampering despite the sticky note reminders, and therefore being completely removed, leaving 0\% untampered sensors.
Deployment 3(B) achieves the highest (100\%) untampered sensors percentage with lowest SNR due to the end user's lack of expert knowledge on the sensor.
Deployment 3(A) balances the untampered sensors percentage and average SNR via LLM-assisted deployment recommendation.
% physical configuration of each sensor unit, such as location and orientation, and compare after the 1 week period. We also record if there were any continuous signal loss due in the networking, or when a sensor unit being unplugged or powered off.
% We take the ratio of the remaining sensors that do not have these anomalies over the total number of sensor units initially deployed.
% Figure \ref{fig:efficiency} shows the relationships between these two metrics for 3 deployments left with the user in a week time span.
% After the week deployment, we find that all 6 sensors deployed in Deployment 2 had been tampered, despite the Sticky Note reminders shown in Figure \ref{dep2}.
% On the other hand, we observed that none have been tampered in Deployment 3(B) and 1 sensor unit was tampered in Deployment 3(A), which were deployed by a family member and assisted with an LLM agent recommendation, respectively.
% This observation models an inverse correlation between percentage of sensor units tampered and the Average SNR, or signal quality.
% However, with the LLM assisted device configuration, we are able to achieve $80\%$ of sensors to remain untampered, while having an average SNR above $25$.
% We verified through these results that the LLM has retained a better balance of user tampering and signal performance than by leaving it solely up for a family member to install or by leaving reminders.

\subsection{Qualitative Analysis for User Experience}

A user study was conducted with participants to understand the overall user experience in the third deployment. We analyzed user experience according to feasibility and acceptance in relation to the participant with cognitive impairment using the system and their adult son who installed and configured the system.

First, to understand the feasibility of our system (Deployment 3(A)), we evaluated whether a family member could interact with the LLM agent in order to generate a recommendation for the sensor configuration.
% Before using the LLM, w
% We pre-generate the sensor and environment schema information before prompting the family member.
We use a parser to pre-generate the text descriptions of the vibration sensor information, like the gain options, constraint to be placed upright on a surface, and the sampling rate, as well as environmental information, such as appliances used in the kitchen, furniture, and health related medication and their locations, all into a structured schema. 
% We first gave the designed parser a text description of the environment, including materials and location of furniture, the shape of the rooms, and orientation of objects.
% The parser fills a schema template with the environment information provided in the description, and was embedded into the prompting of the LLM agent.
% We also embed the instruction prompting, which included on what information is sufficient to reason a recommendation, the reasoning steps to build a relationship map of different fields from the schema to information provided to the user, and using those relationships to generate a recommendation for sensor placement, orientation, and configuration.
% For Deployment 3(A), 
Then, we let a family member ask questions about their deployment to the LLM Agent, for example, ``Is it ok to place the sensor on the kitchen counter''.
The LLM asks for supplementary questions until it reasons that there is sufficient information.
Finally, the LLM agent generates the recommendation for sensor placement, for example, ``directly above or beside its original kitchen counter position".
When asked if the family member could use this recommendation, they agreed that the recommendation was easy to follow, suggesting adequate feasibility.

% As an example, we showcase the conversation between a family member and the LLM about how to configure the sensor system.
% We record and present the initial question and clarifications the user provided the LLM.
% One of the examples, which was to place the sensor in the kitchen, made the family member of Deployment 3 ask this question: ``Is it ok to place the sensor on the kitchen counter''. 
% Due to the insufficient environmental information from the prompt, the LLM asked clarifying questions, such as ``Is the motion sensor facing the rest of the kitchen or pointed toward a specific area (like the sink, stove, or entrance)?''.
% The LLM asked 5 additional questions in total, and only then found it sufficient information to reason with this user provided information.
% After the reasoning, the LLM generates the following recommendation: ``New Coordinates: The sensor should be raised to 1.5 meters (around 5 feet) high on the vertical wall or side of a cabinet, directly above or beside its original kitchen counter position.''
% In this way, we designed a pipeline that can lead a user question about their sensor configuration into a viable solution.

To evaluate feasibility from the participant's perspective, we evaluated whether the participant would be able to live with the sensors in their home without tampering or turning off the system.
We found that of the 2 sets of sensors (10 sensors total), only 1 sensor located in plain sight on the counter was confirmed to be tampered with. Future deployments may consider relocating the sensor to a slightly more obscured location

Finally, the overall acceptability of the system was good both pre- and post-deployment as determined by low reported privacy concerns (rated 1 on a scale from 1 to 5, with 5 being high concern) and affirmative responses to the following questions: ``Security systems for health monitoring and activity tracking could be beneficial'', and  ``The potential benefits of Security Systems in terms of health monitoring and assistance outweigh the potential risks''.

\section{Related Work}\label{sec:related}
%\paragraph{Mentation Challenges: Cognitive Impairments}

\paragraph{IoT for Older Adults Remote Monitoring}
IoT has been used in-home for remote monitoring for older adults, as it is able to improve living standards \cite{Maswadi2020May} and helps track behavior patterns to help diagnose certain neurological conditions \cite{chiuchisan2014approach}.
% However, these IoT devices have not been widely adopted for in-home used, due to the computing requirement, the user experience, and the user concerns.
The computing tasks often include 
% One large requirement for modern home IoT systems is computing the sensor data to predict useful information. Tasks like
human activity recognition \cite{Babangida2022Mar}, fall detection \cite{Yacchirema2018Jan}, and human identification \cite{Rahim2023Aug}.
% , many of which require machine learning models to run.
% Many solutions use cloud computing \cite{Dhirani}, but it comes with ongoing costs for the user.
% Other solutions use edge devices to do limited computing, but this leads to a challenge to fit these models into a memory and computationally constrained device.
On the other hand, user experience plays an important role in the system adoption, which poses obstacles for many of these IoT systems being used for older adults and their family caregivers. 
% Another challenge for in-home IoT systems is user preference.
For example, for user privacy, there are certain types of systems, like vision \cite{Lin2016Jul,Bugeja,Demiris2009Jan} or audio \cite{Zheng2018Nov,Liu2023Sep,Krstulovic2017Sep}, that are largely constrained.
For usability, older adult users also may not want to or forget to wear or maintain wearable devices \cite{Ali2025Mar,VanLaerhoven2015May,Peng2021Jan}, especially older adults with cognitive impairments.
% In order to use privacy preserving modalities, like structural vibration sensors \cite{Pan2019Nov}, many of them require multiple sensor units to coordinate their sensing in different locations, and use the combined signals captured indirectly through the environment for remote monitoring.
% This requirement for managing the throughput and storage of multiple sensor signals, and the increase input size for machine learning, makes privacy aware IoT systems difficult for in-home use.
In this paper, we discuss ways to balance the system performance and user experience with structural vibration based sensing solution and conduct user study with both the older adults and their caregivers.
% manage the throughput and storage of in-home vibration sensor systems and how to lower the computational requirement on activity recognition models to address these hardware and machine learning challenges.

\paragraph{User Experience Evaluation for IoT Systems}
There are many ways to evaluate how people use IoT systems, such as Unified Theory of Acceptance and Use of Technology (UTAUT) \cite{Williams2015Apr},
Privacy Calculus Model \cite{Kim2019Mar}, Human-in-the-Loop Framework \cite{cranor2008framework}, Technology Acceptance Model (TAM) \cite{davis1989technology}, etc. 
We adopt TAM, due to it emphasizes factors influencing continued usage of technology, which is particularly important for assessing long-term remote monitoring systems.
% In order to comprehensively evaluate the user experience and perception of remote monitoring systems, we adapt the 
% Technology Acceptance Model (TAM) \cite{davis1989technology}. 
TAM focuses on 3 components to describe the perception of an IoT system: (1) Perceived Usefulness, or the degree to which a user believes a system will assist them, (2) Perceived Ease of Use, or the degree to which a user believes the system will be free from effort, and (3) Attitude Toward Using, or how willing the user is to using the system in the future.
TAM has been used to model user experience for IoT in retail environments \cite{Patil}, cloud solutions \cite{Robles-Gomez2021Nov}, and healthcare \cite{Alkhwaldi2022Sep}.

\section{Discussion and Future Work} \label{sec:future}
The three iterative real-world deployments allow us to improve the IoT sensing system design and the user interface, which builds solid foundation for larger scale deployments (e.g., 20 homes). We further discuss potential challenges and technical directions next.

\paragraph{Visual Question Answering (VQA) for Precision Recommendation}
The LLM-assisted deployment recommendation we adopt in this paper uses natural language only to obtain information about the sensor, environment, and user preferences, and therefore generate recommendation with limited precision. We will explore using VQA to acquire spatial information more precisely to avoid ambiguity in the recommendation locations while improve the communication efficiency between the user and the LLM. Furthermore, we will explore data representation to allow efficient spatial reasoning. 

\paragraph{Protocol for Long-Term Home Monitoring Evaluation}
In this work, we showcase the labeled data with simulation apartment equipped with high fidelity ground truth collection devices. 
However, in the home deployment, it is difficult if not impossible to acquire such ground truth.
Guiding the participants to conduct a series of activities is one way to collect the ground truth, but it needs to be conducted carefully under the guideline that the process of ground truth collection should not interfere with the data collection.  
For example, the ground truth collection should not include activities that generate vibration data, such as walking around (to observe) or talking loud (to provide activity instruction).
In the future, we will explore interactive ways to label the data during the long term usage through interactive conversation between the user and intelligent systems (e.g., smart speaker). 
The feasibility of doing so with older adults with cognitive impairments need to be verified.
% To enable such data and ground truth gathering under the nature condition, 

\section{Conclusion}\label{sec:conclusion}
In conclusion, we explored challenges of deploying a remote monitoring system in the real world with homes of older adults with cognitive impairment.
We designed a framework, guided by the Geriatric 4Ms (matters most, mentation, mobility, medication) that meets the hardware and software constraints of a home system meant for continuous monitoring, and address the user experience problems with a LLM interface that physically configures the sensor system to meet user needs, tamper-proofing, and privacy concerns. Finally, we analyzed three unique deployments, both in simulations and real world older adult homes, to see the differences in performance and user experience from different iterations of the system.

%%
%% The next two lines define the bibliography style to be used, and
%% the bibliography file.
\bibliographystyle{ACM-Reference-Format}
\bibliography{bib}

@misc{olderPopulation, 
 title={U.S. Older Population Grew From 2010 to 2020 at Fastest Rate Since 1880 to 1890.}, 
 author={Caplan, Zoe},
 url={ https://www.census.gov/library/stories/2023/05/2020-census-united-states-older-population-grew.html},
note = {[Online; accessed 2. Sept. 2025]}, 
year={2023}, 
month={May}
}

@article{mate2021evidence,
  title={Evidence for the 4Ms: interactions and outcomes across the care continuum},
  author={Mate, Kedar and Fulmer, Terry and Pelton, Leslie and Berman, Amy and Bonner, Alice and Huang, Wendy and Zhang, Jinghan},
  journal={Journal of Aging and Health},
  volume={33},
  number={7-8},
  pages={469--481},
  year={2021},
  publisher={Sage Publications Sage CA: Los Angeles, CA}
}

@inproceedings{kimura2024vibrofm,
  title={Vibrofm: Towards micro foundation models for robust multimodal iot sensing},
  author={Kimura, Tomoyoshi and Li, Jinyang and Wang, Tianshi and Chen, Yizhuo and Wang, Ruijie and Kara, Denizhan and Wigness, Maggie and Bhattacharyya, Joydeep and Srivatsa, Mudhakar and Liu, Shengzhong and others},
  booktitle={2024 IEEE 21st International Conference on Mobile Ad-Hoc and Smart Systems (MASS)},
  pages={10--18},
  year={2024},
  organization={IEEE}
}

@article{chang2025leveraging,
  title={Leveraging Audio Representations for Vibration-Based Crowd Monitoring in Stadiums},
  author={Chang, Yen Cheng and Codling, Jesse and Dong, Yiwen and Zhang, Jiale and Chen, Jiasi and Noh, Hae Young and Zhang, Pei},
  journal={arXiv preprint arXiv:2503.17646},
  year={2025}
}

@inproceedings{lee2025rarr,
author = {Lee, Dong Yoon and Weakley, Alyssa and Wei, Hui and Brown, Blake and Carrion, Keyana and Pan, Shijia},
title = {RRAR: Robust Real-World Activity Recognition with Vibration by Scavenging Near-Surface Audio Online},
year = {2025},
isbn = {9798400719783},
publisher = {Association for Computing Machinery},
address = {New York, NY, USA},
url = {https://doi.org/10.1145/3737901.3768365},
doi = {10.1145/3737901.3768365},
booktitle = {Proceedings of the 3rd ACM International Workshop on Intelligent Acoustic Systems and Applications},
pages = {1–6},
numpages = {6},
location = {Hong Kong, China},
series = {IASA '25}
}

@misc{olderPopulation2, 
 title={Demographic Turning Points for the United States: Population Projections for 2020 to 2060. }, 
 author={Vespa, Jonathan and  Medina, Lauren and Armstrong, David},
 url={ https://www.census.gov/content/dam/Census/library/publications/2020/demo/p25-1144.pdf},
note = {[Online; accessed 2. Sept. 2025]}, 
year={ 2020}, 
month={February}
}

@misc{olderPopulation4, 
 title={Families sacrifice time and money caring for loved ones. A new caregiver calculator reveals the high cost. }, 
 author={Quraishi, Ash-har and Zalani, Aparna and Beard, Ryan and Munoz, Aaron},
 url={https://www.cbsnews.com/news/caregiver-calculator-cost-for-families/},
note = {[Online; accessed 2. Sept. 2025]}, 
year={ 2024}, 
month={October}
}

@misc{G4Ms, 
 title={The 4Ms Explained}, 
 author={Age-Friendly Care},
 url={https://www.agefriendlycare.psu.edu/the-4ms-explained},
note = {[Online; accessed 11. Nov. 2025]}
}

@article{bai2018empirical,
  title={An empirical evaluation of generic convolutional and recurrent networks for sequence modeling},
  author={Bai, Shaojie and Kolter, J Zico and Koltun, Vladlen},
  journal={arXiv preprint arXiv:1803.01271},
  year={2018}
}

@incollection{Zhou,
	author = {Zhou, Zhongna and Dai, Wenqing and Eggert, Jay and Giger, Jarod T. and Keller, James and Rantz, Marilyn},
	title = {{A real-time system for in-home activity monitoring of elders}},
	booktitle = {{2009 Annual International Conference of the IEEE Engineering in Medicine and Biology Society}},
	pages = {03--06},
	isbn = {978-1-4244-3296-7},
	publisher = {IEEE},
	doi = {10.1109/IEMBS.2009.5334915}
}

@article{Muller2017Sep,
	author = {M{\ifmmode\ddot{u}\else\"{u}\fi}ller, Andre Matthias and Blandford, Ann and Yardley, Lucy},
	title = {{The conceptualization of a Just-In-Time Adaptive Intervention (JITAI) for the reduction of sedentary behavior in older adults}},
	journal = {mHealth},
	volume = {3},
	pages = {37},
	year = {2017},
	month = sep,
	doi = {10.21037/mhealth.2017.08.05}
}

@article{Leslie2021Mar,
	author = {Leslie, Myles and Gray, Robin Patricia and Khayatzadeh-Mahani, Akram},
	title = {{What is {`}care quality{'} and can it be improved by information and communication technology? A typology of family caregivers' perspectives}},
	journal = {Scand. J. Caring Sci.},
	volume = {35},
	number = {1},
	pages = {220--232},
	year = {2021},
	month = mar,
	issn = {0283-9318},
	publisher = {John Wiley {\&} Sons, Ltd},
	doi = {10.1111/scs.12837}
}

@incollection{Singh2019Oct,
	author = {Singh, Akash Deep and Sandha, Sandeep Singh and Garcia, Luis and Srivastava, Mani},
	title = {{RadHAR: Human Activity Recognition from Point Clouds Generated through a Millimeter-wave Radar}},
	booktitle = {{ACM Conferences}},
	pages = {51--56},
	year = {2019},
	month = oct,
	publisher = {Association for Computing Machinery},
	address = {New York, NY, USA},
	doi = {10.1145/3349624.3356768}
}

@incollection{Gordon,
	author = {Gordon, Dawud and Schmidtke, Hedda Rahel and Beigl, Michael and von Zengen, Georg},
	title = {{A novel micro-vibration sensor for activity recognition: Potential and limitations}},
	booktitle = {{International Symposium on Wearable Computers (ISWC) 2010}},
	journal = {Published in: International Symposium on Wearable Computers (ISWC) 2010},
	pages = {10--13},
	publisher = {IEEE},
	doi = {10.1109/ISWC.2010.5665861}
}

@incollection{Kashimoto,
	author = {Kashimoto, Yukitoshi and Fujiwara, Masashi and Fujimoto, Manato and Suwa, Hirohiko and Arakawa, Yutaka and Yasumoto, Keiichi},
	title = {{ALPAS: Analog-PIR-Sensor-Based Activity Recognition System in Smarthome}},
	booktitle = {{2017 IEEE 31st International Conference on Advanced Information Networking and Applications (AINA)}},
	journal = {Published in: 2017 IEEE 31st International Conference on Advanced Information Networking and Applications (AINA)},
	pages = {27--29},
	issn = {1550-445X},
	publisher = {IEEE},
	doi = {10.1109/AINA.2017.33}
}

@article{Lara2012Nov,
	author = {Lara, Oscar D. and Labrador, Miguel A.},
	title = {{A Survey on Human Activity Recognition using Wearable Sensors}},
	journal = {IEEE Commun. Surv. Tutorials},
	volume = {15},
	number = {3},
	pages = {1192--1209},
	year = {2012},
	month = nov,
	publisher = {IEEE},
	doi = {10.1109/SURV.2012.110112.00192}
}

@article{Chen2018Mar,
	author = {Chen, Yi and Yu, Li and Ota, Kaoru and Dong, Mianxiong},
	title = {{Robust Activity Recognition for Aging Society}},
	journal = {IEEE J. Biomed. Health Inf.},
	volume = {22},
	number = {6},
	pages = {1754--1764},
	year = {2018},
	month = mar,
	publisher = {IEEE},
	doi = {10.1109/JBHI.2018.2819182}
}

@article{Shi2022Jul,
	author = {Shi, Zhenguo and Cheng, Qingqing and Zhang, J. Andrew and Da Xu, Richard Yi},
	title = {{Environment-Robust WiFi-Based Human Activity Recognition Using Enhanced CSI and Deep Learning}},
	journal = {IEEE IoT J.},
	volume = {9},
	number = {24},
	pages = {24643--24654},
	year = {2022},
	month = jul,
	publisher = {IEEE},
	doi = {10.1109/JIOT.2022.3192973}
}

@incollection{Patil,
	author = {Patil, Kanchan},
	title = {{Retail adoption of Internet of Things: Applying TAM model}},
	booktitle = {{2016 International Conference on Computing, Analytics and Security Trends (CAST)}},
	journal = {Published in: 2016 International Conference on Computing, Analytics and Security Trends (CAST)},
	pages = {19--21},
	publisher = {IEEE},
	doi = {10.1109/CAST.2016.7915003}
}

@article{Robles-Gomez2021Nov,
	author = {Robles-G{\ifmmode\acute{o}\else\'{o}\fi}mez, Antonio and Tobarra, Llanos and Pastor-Vargas, Rafael and Hern{\ifmmode\acute{a}\else\'{a}\fi}ndez, Roberto and Haut, Juan M.},
	title = {{Analyzing the Users{'} Acceptance of an IoT Cloud Platform Using the UTAUT/TAM Model}},
	journal = {IEEE Access},
	volume = {9},
	pages = {150004--150020},
	year = {2021},
	month = nov,
	issn = {2169-3536},
	publisher = {IEEE},
	doi = {10.1109/ACCESS.2021.3125497}
}

@incollection{Alkhwaldi2022Sep,
	author = {Alkhwaldi, Abeer F. and Abdulmuhsin, Amir A.},
	title = {{Understanding User Acceptance of IoT Based Healthcare in Jordan: Integration of the TTF and TAM}},
	booktitle = {{Digital Economy, Business Analytics, and Big Data Analytics Applications}},
	journal = {SpringerLink},
	pages = {191--213},
	year = {2022},
	month = sep,
	issn = {1860-9503},
	isbn = {978-3-031-05258-3},
	publisher = {Springer},
	address = {Cham, Switzerland},
	doi = {10.1007/978-3-031-05258-3_17}
}

@article{Yu2023Oct,
	author = {Yu, Zihan and He, Liang and Wu, Zhen and Dai, Xinyu and Chen, Jiajun},
	title = {{Towards Better Chain-of-Thought Prompting Strategies: A Survey}},
	journal = {arXiv},
	year = {2023},
	month = oct,
	eprint = {2310.04959},
	doi = {10.48550/arXiv.2310.04959}
}

@article{Maswadi2020May,
	author = {Maswadi, Kholoud and Ghani, Norjihan Binti Abdul and Hamid, Suraya Binti},
	title = {{Systematic Literature Review of Smart Home Monitoring Technologies Based on IoT for the Elderly}},
	journal = {IEEE Access},
	volume = {8},
	pages = {92244--92261},
	year = {2020},
	month = may,
	issn = {2169-3536},
	publisher = {IEEE},
	doi = {10.1109/ACCESS.2020.2992727}
}

@article{chiuchisan2014approach,
  title={An approach of a decision support and home monitoring system for patients with neurological disorders using internet of things concepts},
  author={Chiuchisan, IULIANA and Geman, OANA},
  journal={WSEAS Transactions on Systems},
  volume={13},
  number={1},
  pages={460--469},
  year={2014},
  publisher={WSEAS}
}

@article{Babangida2022Mar,
	author = {Babangida, Lawal and Perumal, Thinagaran and Mustapha, Norwati and Yaakob, Razali},
	title = {{Internet of Things (IoT) Based Activity Recognition Strategies in Smart Homes: A Review}},
	journal = {IEEE Sens. J.},
	volume = {22},
	number = {9},
	pages = {8327--8336},
	year = {2022},
	month = mar,
	publisher = {IEEE},
	doi = {10.1109/JSEN.2022.3161797}
}

@article{davis1989technology,
  title={Technology acceptance model: TAM},
  author={Davis, Fred D and others},
  journal={Al-Suqri, MN, Al-Aufi, AS: Information Seeking Behavior and Technology Adoption},
  volume={205},
  number={219},
  pages={5},
  year={1989}
}

@article{Rahim2023Aug,
	author = {Rahim, Asif and Zhong, Yanru and Ahmad, Tariq and Ahmad, Sadique and P{\l}awiak, Pawe{\l} and Hammad, Mohamed},
	title = {{Enhancing Smart Home Security: Anomaly Detection and Face Recognition in Smart Home IoT Devices Using Logit-Boosted CNN Models}},
	journal = {Sensors},
	volume = {23},
	number = {15},
	pages = {6979},
	year = {2023},
	month = aug,
	issn = {1424-8220},
	publisher = {Multidisciplinary Digital Publishing Institute},
	doi = {10.3390/s23156979}
}

@article{Lin2016Jul,
	author = {Lin, Huichen and Bergmann, Neil W.},
	title = {{IoT Privacy and Security Challenges for Smart Home Environments}},
	journal = {Information},
	volume = {7},
	number = {3},
	pages = {44},
	year = {2016},
	month = jul,
	issn = {2078-2489},
	publisher = {Multidisciplinary Digital Publishing Institute},
	doi = {10.3390/info7030044}
}

@article{Zheng2018Nov,
	author = {Zheng, Serena and Apthorpe, Noah and Chetty, Marshini and Feamster, Nick},
	title = {{User Perceptions of Smart Home IoT Privacy}},
	journal = {Proc. ACM Hum.-Comput. Interact.},
	volume = {2},
	number = {CSCW},
	pages = {1--20},
	year = {2018},
	month = nov,
	publisher = {Association for Computing Machinery},
	doi = {10.1145/3274469}
}

@article{Ali2025Mar,
	author = {Ali, Amir and Montanaro, Teodoro and Sergi, Ilaria and Carrisi, Simone and Galli, Daniele and Distante, Cosimo and Patrono, Luigi},
	title = {{An Innovative IoT and Edge Intelligence Framework for Monitoring Elderly People Using Anomaly Detection on Data from Non-Wearable Sensors}},
	journal = {Sensors},
	volume = {25},
	number = {6},
	pages = {1735},
	year = {2025},
	month = mar,
	issn = {1424-8220},
	publisher = {Multidisciplinary Digital Publishing Institute},
	doi = {10.3390/s25061735}
}

@article{Williams2015Apr,
	author = {Williams, Michael D. and Rana, Nripendra P. and Dwivedi, Yogesh K.},
	title = {{The unified theory of acceptance and use of technology (UTAUT): a literature review}},
	journal = {Journal of Enterprise Information Management},
	volume = {28},
	number = {3},
	pages = {443--488},
	year = {2015},
	month = apr,
	issn = {1741-0398},
	publisher = {Emerald Publishing},
	doi = {10.1108/JEIM-09-2014-0088}
}

@article{Kim2019Mar,
	author = {Kim, Dongyeon and Park, Kyuhong and Park, Yongjin and Ahn, Jae-Hyeon},
	title = {{Willingness to provide personal information: Perspective of privacy calculus in IoT services}},
	journal = {Computers in Human Behavior},
	volume = {92},
	pages = {273--281},
	year = {2019},
	month = mar,
	issn = {0747-5632},
	publisher = {Pergamon},
	doi = {10.1016/j.chb.2018.11.022}
}

@article{cranor2008framework,
  title={A framework for reasoning about the human in the loop},
  author={Cranor, Lorrie F},
  year={2008},
  publisher={Carnegie Mellon University}
}

@incollection{Bugeja,
	author = {Bugeja, Joseph and Jacobsson, Andreas and Davidsson, Paul},
	title = {{On Privacy and Security Challenges in Smart Connected Homes}},
	booktitle = {{2016 European Intelligence and Security Informatics Conference (EISIC)}},
	journal = {Published in: 2016 European Intelligence and Security Informatics Conference (EISIC)},
	pages = {17--19},
	publisher = {IEEE},
	doi = {10.1109/EISIC.2016.044}
}

@article{Demiris2009Jan,
	author = {Demiris, George and Oliver, Debra Parker and Giger, Jarod and Skubic, Marjorie and Rantz, Marilyn},
	title = {{Older adults' privacy considerations for vision based recognition methods of eldercare applications}},
	journal = {Technol. Health Care},
	volume = {17},
	number = {1},
	pages = {41--48},
	year = {2009},
	month = jan,
	issn = {0928-7329},
	publisher = {SAGE Publications},
	doi = {10.3233/THC-2009-0530}
}

@article{Liu2023Sep,
	author = {Liu, Yuchen and Kapadia, Apu and Williamson, Donald},
	title = {{Privacy-preserving and Privacy-attacking Approaches for Speech and Audio -- A Survey}},
	journal = {arXiv},
	year = {2023},
	month = sep,
	eprint = {2309.15087},
	doi = {10.48550/arXiv.2309.15087}
}

@incollection{Krstulovic2017Sep,
	author = {Krstulovi{\ifmmode\acute{c}\else\'{c}\fi}, Sacha},
	title = {{Audio Event Recognition in the Smart Home}},
	booktitle = {{Computational Analysis of Sound Scenes and Events}},
	journal = {SpringerLink},
	pages = {335--371},
	year = {2017},
	month = sep,
	isbn = {978-3-319-63450-0},
	publisher = {Springer},
	address = {Cham, Switzerland},
	doi = {10.1007/978-3-319-63450-0_12}
}

@article{VanLaerhoven2015May,
	author = {Van Laerhoven, Kristof and Borazio, Marko and Burdinski, Jan Hendrik},
	title = {{Wear is Your Mobile? Investigating Phone Carrying and Use Habits with a Wearable Device}},
	journal = {Front. ICT},
	volume = {2},
	pages = {141865},
	year = {2015},
	month = may,
	issn = {2297-198X},
	publisher = {Frontiers},
	doi = {10.3389/fict.2015.00010}
}

@article{Peng2021Jan,
	author = {Peng, Wei and Li, Lin and Kononova, Anastasia and Cotten, Shelia and Kamp, Kendra and Bowen, Marie},
	title = {{Habit Formation in Wearable Activity Tracker Use Among Older Adults: Qualitative Study}},
	journal = {JMIR mHealth and uHealth},
	volume = {9},
	number = {1},
	pages = {e22488},
	year = {2021},
	month = jan,
	publisher = {JMIR Publications Inc., Toronto, Canada},
	doi = {10.2196/22488}
}

@article{Yacchirema2018Jan,
	author = {Yacchirema, Diana and de Puga, Jara Su{\ifmmode\acute{a}\else\'{a}\fi}rez and Palau, Carlos and Esteve, Manuel},
	title = {{Fall detection system for elderly people using IoT and Big Data}},
	journal = {Procedia Comput. Sci.},
	volume = {130},
	pages = {603--610},
	year = {2018},
	month = jan,
	issn = {1877-0509},
	publisher = {Elsevier},
	doi = {10.1016/j.procs.2018.04.110}
}

%%
%% If your work has an appendix, this is the place to put it.
\appendix

\end{document}